# Mpemba Paradox Revisited: Numerical Reinforcement

Xi Zhang[1,2,*] Yongli Huang[3,*], Zengsheng Ma[3], Chang Q Sun[1-3, #]

Preface

Inspired by responses to the work ArXived (http://arxiv.org/abs/1310.6514), the authors solved the one-dimensional, nonlinear Fourier heat transportation initial and boundary condition problem using the finite element method. Examined all possible parameters, results reinforced our previous suggestions: three key factors: heat emission, conduction, and dissipation dictates this "source-path-drain" problem.

The revisit reveals the following:

- **Hydrogen bond has memory effect** to emit energy at a rate, or with a relaxation time, depending **on initial energy storage**.
- **Skin supersolidity** creates gradients in density, specific heat, thermal conductivity, and temperature field for heat conduction in liquid with the critical thermal diffusion coefficient ratio of $\alpha_{Skin}/\alpha_{bulk} \geq \rho_{Bulk}/\rho_{Skin} = 4/3$ (optimal at 1.48).
- **Convection** alone produces no Mpemba effect at all though this explanation is most popular.
- **Sensitivity examination** revealed that Mpemba effect happens only in the **highly non-diabetic "source-path-drain"** cycling system.

Our recent work laid firm foundations towards the solution to this project:

## Abstract

**Non-linear Fourier's Law solution reinforced that the Mpemba paradox arises *intrinsically* from the anomalous relaxation of the hydrogen bond (O:H-O) that exhibits memory effect and forms eddy current in liquid. Firstly, heating stores energy into water by stretching the intermolecular O:H nonbond and shortening the intermolecular H-O bond via a Coulomb coupling mechanism; cooling does oppositely, like releasing a highly deformed bungee, to emit heat at a rate that depends on the initial energy storage. Secondly, heating and molecular undercoordination in the skin elevate jointly gradients of mass density, specific heat, and thermal conductivity in the liquid, favoring heat flowing towards the skin. Convection alone raises the skin temperature but creates no Mpemba effect. Being sensitive to the liquid volume, Mpemba effect proceeds only in the non-adiabatic source-drain interface ambient with a characteristic relaxation time that drops exponentially with the rise of the initial source temperature.**

Mpemba paradox [3], i.e., warmer water freezes faster than colder water does, has baffled thinkers like Francis Bacon and René Descartes dating back to Aristotle [4]. As firstly noted by Aristotle [4]: "The fact that the water has previously been warmed contributes to its freezing quickly: for so it cools sooner". Hence many people, when they want to cool water quickly, begin by putting it in the sun.

Although there is anecdotal support for this paradox [9], no agreement has been made on exactly what the effect is and under what circumstances it occurs [11, 12]. This phenomenon remains a paradox in thermodynamics albeit so many possible explanations in terms of evaporation [14], frost [15], solutes [16], convection [17, 18], supercooling [15], etc [19]. According to the winner [20] of a competition held in 2012 by the Royal Society of Chemistry calling for papers offering explanations to the Mpemba paradox, the effect of convection enhances the probability of warmer water freezing. It was best explained that [20] the flow in the first stage of cooling from $\theta_h$ (for hot) to $\theta_c$ (for cool) continues throughout the rest cooling process, accelerating cooling of the initially hotter water even after it reaches $\theta_c$.

Despite claims often made by one source or another, there is no well-agreed explanation for how this phenomenon occurs. Actually, up to now explanations has just been guesswork about what is happening during the Mpemba effect with focusing mainly on extrinsic factors. Little attention [21, 22] has yet been paid to the nature of the heat source or mechanism behind the entire "source-path-drain" cycling system. This Letter shows evidence that this effect arises intrinsically from the anomalous relaxation of the hydrogen bond due to thermal [23] and molecular undercoordination [5] effects.

**Figure 1** illustrates the relaxation dynamics of the O:H-O bond in water. The O:H-O bond is composed of the O:H and the H-O part other than either of them alone [5]. The intermolecular O:H van der Waals (vdW) bond forms with lone pair interaction energy at the $10^{-2}$ eV level and the intramolecular H-O covalent bond with exchange interaction energy around 4.0 eV. This segmented O:H-O bond forms a pair of asymmetric, H-bridged oscillators coupled by Coulomb repulsion between electron pairs on adjacent oxygen atoms, with H atom being the coordination origin [2, 13].

Generally, heating raises the energy of a substance by lengthening and softening all bonds involved; cooling does oppositely with volume contraction and heat emission. However, the O:H nonbond in water follows actively the regular rule of thermal expansion because of its relatively lower specific heat; the H-O part acts oppositely as a slave to relax in the same direction but by different amount [23].

The heating-cooling cycling reverses oxygen coordinates along the potential paths of **Figure 1**. The red spheres represent oxygen atoms in the hot state and the blue ones in the cold state. Heating stores energy into the liquid by deforming the O:H nonbond and the H-O bond. At cooling, the shorter and stiffer H-O bond will be kicked up along its potential path by O:H nonbond cooling contraction and the repulsive force between O atoms, which ejects energy to the drain through conduction and dissipation. This cooling process is like suddenly releasing a deformed bungee from different extents of deformation with the kicking by O:H contraction as an additional force propelling the energy emission. The rate of energy ejection, or thermal momentum, is expected to be dependent on the initial energy storage.

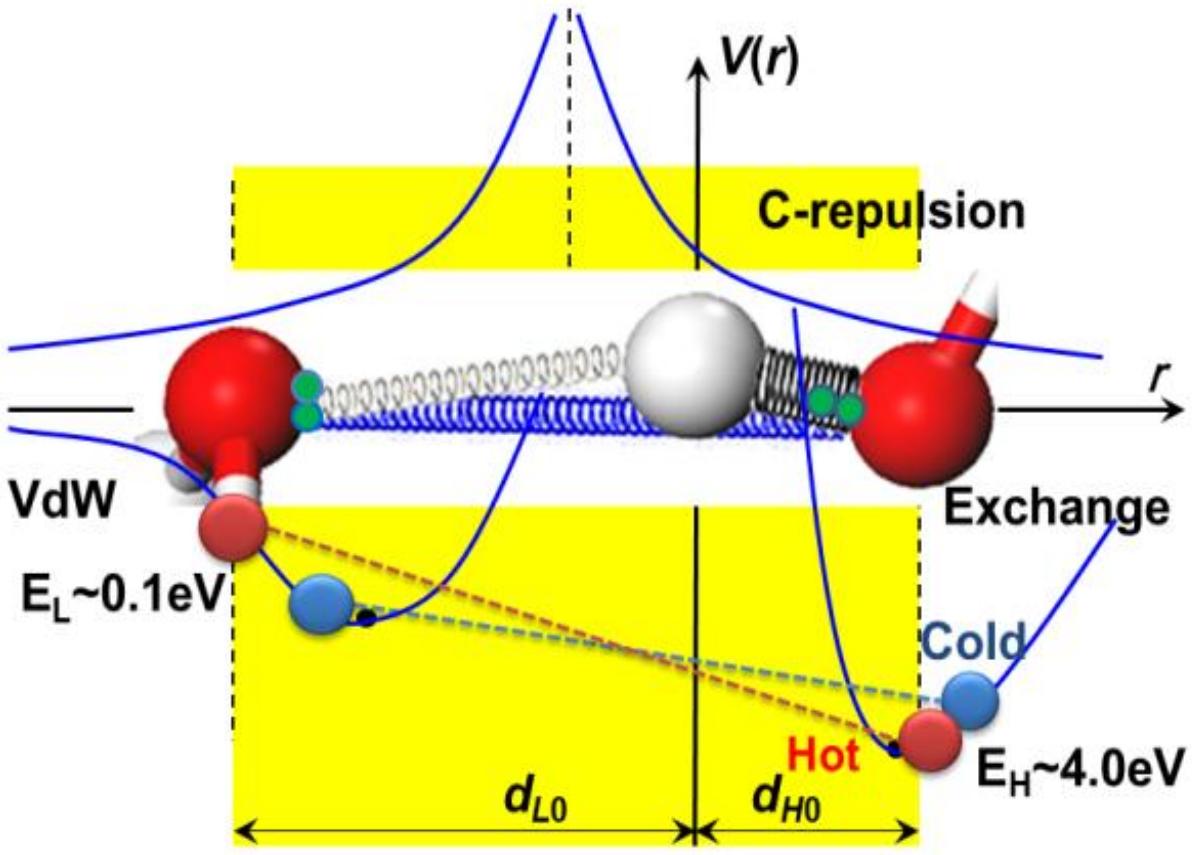

**Figure 1 O:H-O bond short-range interactions and cooperative relaxation.** The O:H-O bond is composed of the weak O:H nonbond in the left-hand side with van der Waals interaction (~0.1 eV) and the strong H-O covalent bond in the right with exchange interaction (~4.0 eV) with H atom as the coordination origin. Inter-electron-pair (small paring dots on oxygen) Coulomb repulsion couples these two parts to relax in the same direction but by different amounts under applied stimuli such as heating (red spheres) or cooling (blue spheres) associated with energy change along the respective potential path. Heating and undercoordination enhance each other in the process of relaxation.

In the process of molecular undercoordination, H-O contraction dominates because of the bond-order-length-strength correlation [5]. Measurements show that heating and molecular undercoordination share the same attribute of H-O contraction and O:H elongation and the associated blueshift of the H-O bond energy $E_H$ and the vibration frequency $\omega_H$ (see Fig S1-2 in Supplementary Information)[24, 25]. For examples, heating from 0 °C to 90 °C raises the $\omega_H$ from 3190 to 3260 $cm^{-1}$ [6] and the O 1s binding energy by 0.15 eV [26]. The $\omega_H$ increases[27] from 3200 to 3450 $cm^{-1}$ and the O 1s energy increases by 1.5 eV [28, 29] if one moves from the liquid bulk to the skin. The O---O distance increases by some 10% from the bulk value of 2.695[2] ~ 2.700 Å [30] to 2.960 Å [31] in the skin, which is accompanied [2] by the $d_H$ shortening from 1.00 to 0.84 Å and a density drop from the ideal value of 1.0 at 4 °C to 0.75 $g\cdot cm^{-3}$.

Water molecules with fewer than four neighbors form a supersolid phase that is elastic [27], polarized [32-34], hydrophobic [35, 36], viscostic [37], thermally stable [38] with density being lower than that of ice (0.92 $g\cdot cm^{-3}$)[5]. The integral of the specific heat of the H-O bond, $\eta_H$, over the full temperature range is proportional to the cohesive energy $E_H$ of the H-O bond [23] so the $\eta_H$ varies with the thermal slope of the $E_H$. The thermal conductivity $\kappa \approx \eta_H v l \approx \eta_H (C_p)$ correlates to the specific heat $\eta_H$ or its equivalence at constant pressure with $v$ and $l$ being the velocity and the mean-free-path of phonons, respectively. Therefore, variation of the $dE_H / d\theta$ and the mass density [2] in the skin region redefine the local diffusion coefficient $\alpha = \kappa/(\rho C_p)$ in the Fourier equation, which heat conduction in the liquid follows. The addition of the supersolid skin creates the thermal conductivity gradient, which is expected to ensure thermal eddy current to flow outwardly towards the skin.

To verify the prediction on the presence of thermal conductivity gradient due to the joint effect of heating and skin supersolidity, we solved the one-dimensional nonlinear Fourier equation [39] numerically by introducing the supersolid skin [5] in a tube container. Water in a cylindrical tube can be divided into the bulk (B) and the skin (S) region along the x-axial direction and put the tube into a drain of constant temperature 0 °C. The other end is open to the drain without the skin. The heat transfer in the partitioned fluid follows this eequation:

$$\frac{\partial\theta}{\partial t} = \alpha_j(\theta)\nabla^2\theta - \mathbf{u}\cdot\frac{\partial\theta}{\partial x};(j = Skin\ or\ Bulk)$$
$$\alpha(\theta) = \frac{\kappa(\theta)}{\rho(\theta)C_p(\theta)}, \tag{1}$$

The first term describes diffusion and the second convection. The known temperature dependence of the thermal conductivity $\kappa_B(\theta)$, mass density $\rho_B$ $(\theta)$, and specific heat under constant pressure $C_{pB}(\theta)$, determines the diffusion coefficient $\alpha_B$. For simplicity, one can take the $\alpha_S$ in the skin (z = 2) as an adjustable for optimization. $\mathbf{u} = 10^{-4}$ m/s is the constant velocity field. The following initial and boundary conditions apply:

$$\theta(t=0) = \theta_i$$
$$\begin{cases} \theta + h\dfrac{\partial\theta}{\partial x} = 0\ ^\circ C & (x = -L_0, end\ of\ bulk) \\ \theta(0^-) = \theta(0^+) & (x = 0, skin-bulk\ interface) \\ \theta - h\dfrac{\partial\theta}{\partial x} = 0\ ^\circ C & (x = L_1, end\ of\ skin) \end{cases}$$

The heat transfer coefficient *h* of the cooling ends remains at 30 w/($m^2$K) and $\mathbf{u} = 10^{-4}$ m/s is the heat flow convection velocity in bulk water [40].

The partial differential equation is solved using finite element method. In numerical calculations, computer reads in the digitized data of $\rho(\theta)$, $\kappa(\theta)$, and $C_p(\theta)$ for the $\alpha_B(\theta)$. The skin $\alpha_S(\theta)$ is obtained by multiplying $\alpha_B(\theta)$ with a coefficient that is 4/3 or greater due to molecular undercoordination. Figure 2 and Figure 3 compare the examination results of the following parameters.

| Parameters | | Observations | |
|---|---|---|---|
| Convection flow rate v (m/s) | Thermal diffusion coefficient $\alpha_S/\alpha_B$ | Mpemba effect | Skin-bottom $\Delta\theta$ |
| 0 | 1 | No | No |
| $10^{-4}$ | | | yes |
| 0 | 1.48 | Yes | Negligible |
| $10^{-4}$ | | | Match measurements |

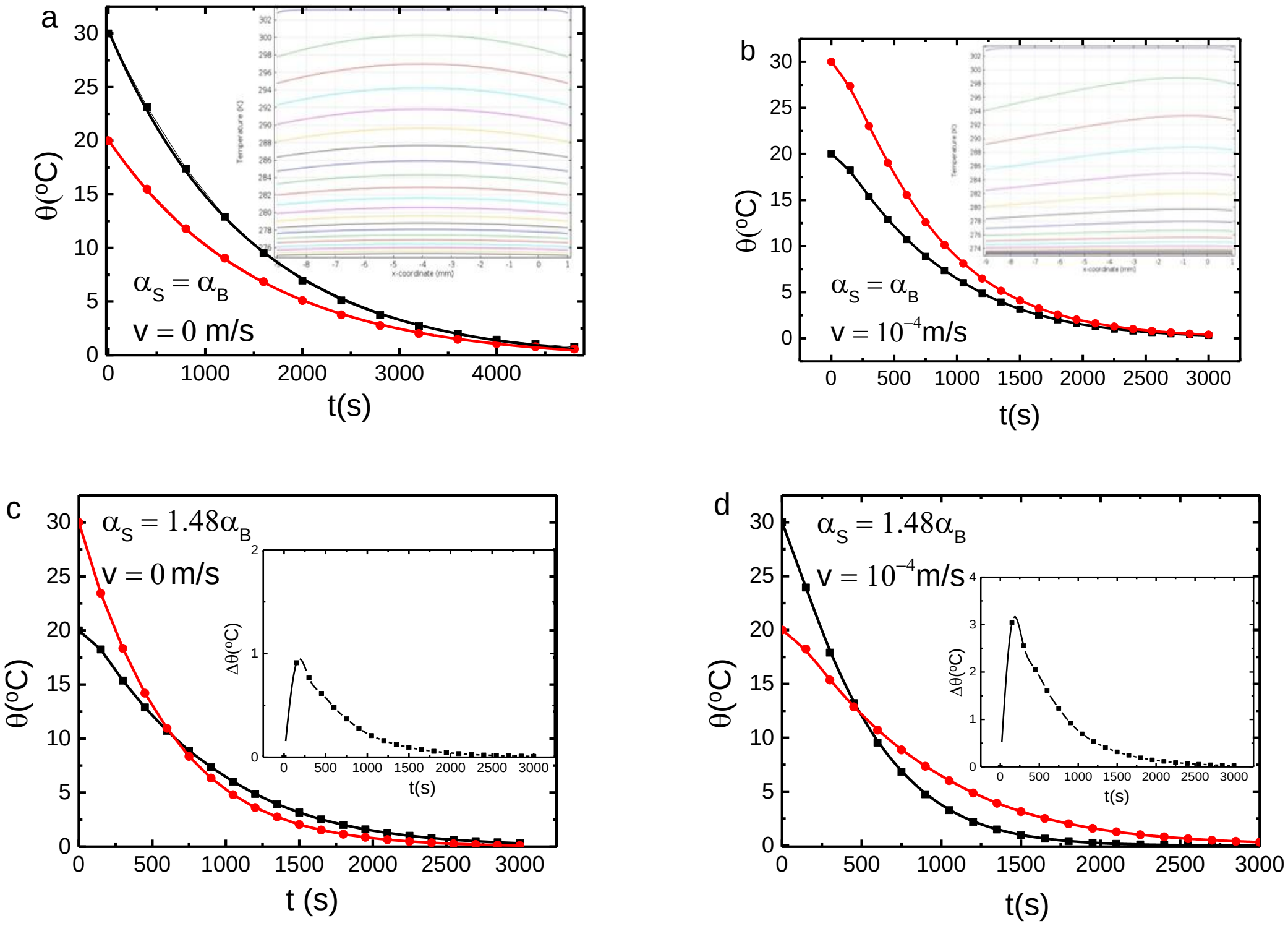

Figure 2 θ(t) profiles for $\theta_i$ = 20 and 30 °C with the given conditions (convection heat flow velocity v and thermal diffusion coefficient ratio $\alpha_S/\alpha_B$) with insets the θ(x, t) profiles for $\theta_i$ = 30 °C in (a) and (b) and the Δθ(t) profile for $\theta_i$ = 20 and 30 °C in (c and d). Convection (v≠ 0) raises only the skin temperature by a maxima of Δθ(t) = 4 °C without Mpemba effect. Skin supersolidity ($\alpha_S/\alpha_B$ = 1.48) creates Mpemba effect with negligible skin temperature elevation. (d) Convection supplements the skin supersolidity to create Mpemba effect raising the skin temperature by a maximal of Δθ(t) = 3 °C.

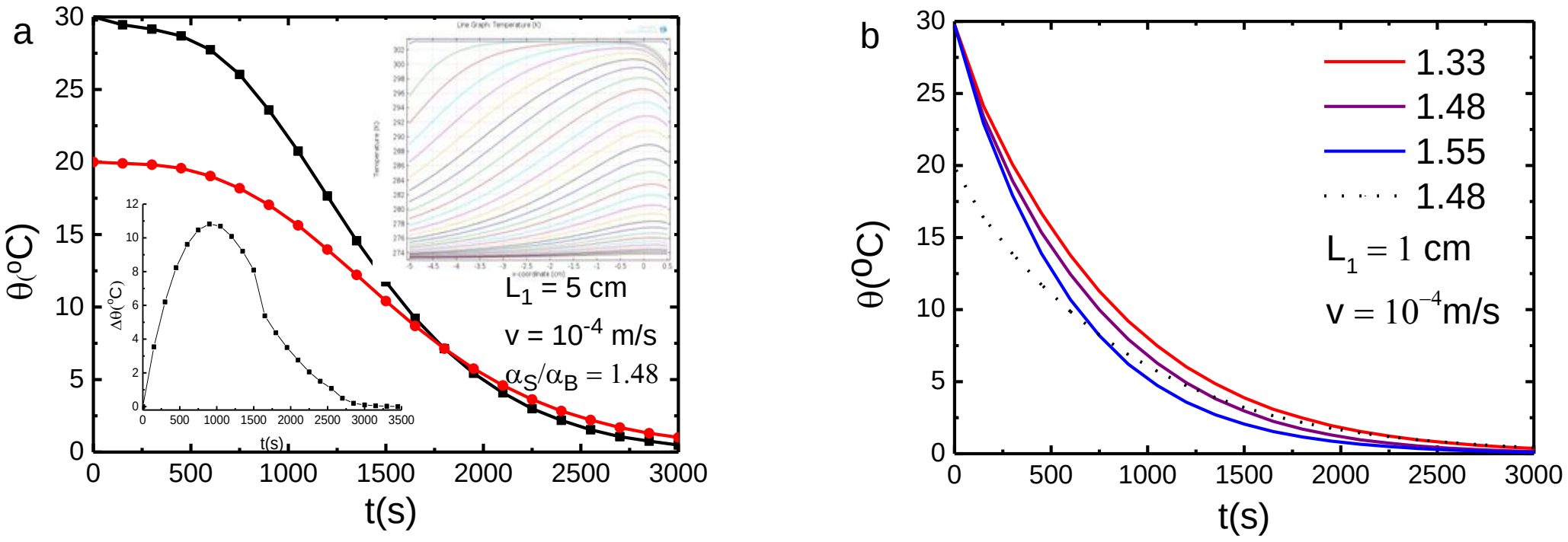

Figure 3 Sensitivity to (a) tube size and (b) $\alpha_S/\alpha_B$ ratio. (a) Volume inflation prolongs the time to reach the crossing temperature and raises the skin temperature in the present one-dimensional system. Continuing liquid volume inflation may annihilate the Mpemba effect because of the transiting from the non-adiabatic to adiabatic ambient, particularly in a three-

dimensional system. (b) The critical $\alpha_S/\alpha_B \geq 4/3$ corresponds to the ratio of $\rho_B/\rho_S = 1/0.75$. An increase of the $\alpha_S/\alpha_B$ shortens the time to the crossing point.

Numerical search based on the known temperature dependence of the thermal diffusion coefficient $\alpha(\theta)$ and the skin effect revealed the following: i) the skin supersolidity enables the Mpemba effect; ii) heat convection alone only raises the skin temperature; iii) the critical ratio of the thermal diffusion coefficient $\alpha_S/\alpha_B \geq \rho_B/\rho_S = 4/3$;[2] and, iv) the crossing temperature of different $\theta(t)$ curves is sensitive to the tube length or volume of the liquid in the one-dimension tube cell.

The optimal results in **Figure 4** match what Bregović, Mpemba, and Osborne have observed [3, 20]: i) hotter water freezes faster than colder water under the same conditions; ii) the temperature $\theta$ drops exponentially with cooling time (t) for transiting water into ice; iii) the skin is warmer than sites near the bottom in a beaker of water being cooled; and, iv) the skin of hotter water is even warmer throughout the course of cooling[3], which indicates that the heat convection and diffusion rate increase with temperature.

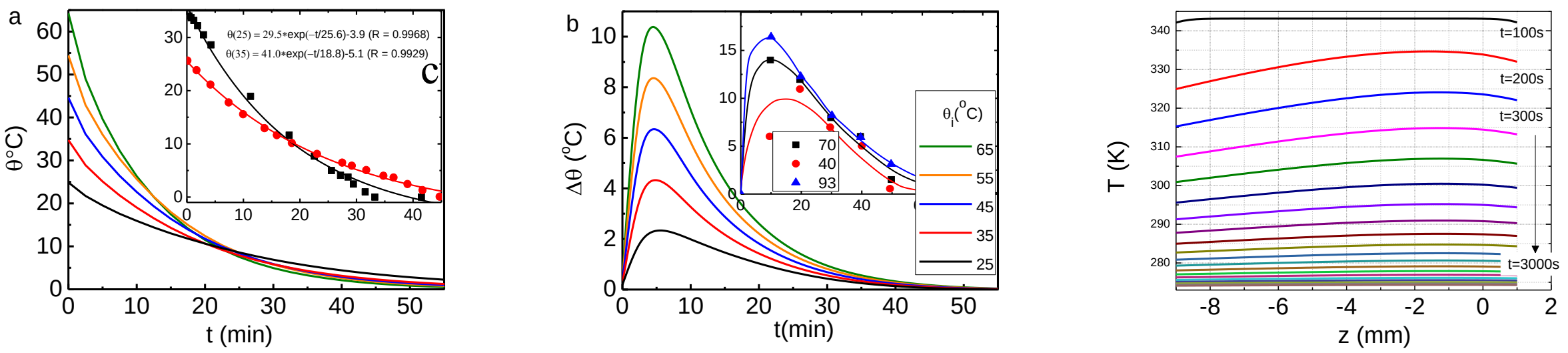

**Figure 4** Theoretical reproduction of (a) the $\theta(t)$ and (b) the skin-bottom $\Delta\theta(t)$ curves for water cooling from different $\theta_i$ as recorded by Bregović, Mpemba and Osborne (insets)[3, 20]. (c) The $\theta(x, t)$ profiles under the optimal conditions: $\alpha_S = 1.48\alpha_B$, tube length 10 mm, skin thickness 2 nm. Formulating the experimental curves shows the $\theta_i$ dependent of the relaxation time $\tau$.

Next, we prove that the O:H-O bond possesses memory whose relaxation rate depends on initial energy storage. The following formulates the $\theta(t)$ decay profiles in **Figure 4**a,

$$\begin{cases} d\theta = -\tau_i^{-1}\theta dt & (\textit{decay function}) \\ \tau_i^{-1} = \sum_j \tau_{ji}^{-1} & (\textit{relaxation time}) \end{cases} \tag{2}$$

The $\tau_i$ is the sum of $\tau_{ji}$ over all the possible jth processes of heat loss during cooling. Conducting experiments under identical conditions is necessary to minimize artifacts arising from processes such as radiation, source/drain volume ratio, exposure area, container material, etc. For instance, cooling one drop of 1 mL water needs shorter time than cooling one cup of 200 mL water at the same $\theta_i$ under the same conditions.

The measured θ(t) curves [3, 20] in **Figure 4** provides the $d\theta/dt=-\tau_i^{-1}\theta$. Converting the density ρ(θ) profile into the $d_H(\theta)=1.0042\text{-}2.7912\times10^{-5}\exp\left[(\theta+273)/57.2887\right]$ (Å) formulates the temperature dependence of the H-O bond length [2]. The $E_H = 0.5k_H(\Delta d_H)^2$ approximates the energy stored in the H-O bond. Thus, one can determine the instantaneous velocity of the H-O bond relaxation in length and energy:

$$\begin{cases} \dfrac{d\left(d_H(\theta)\right)}{dt} & = \dfrac{\partial\left(d_H(\theta)\right)}{\partial\theta}\dfrac{d\theta}{dt} = -\tau_i^{-1}\theta\dfrac{\Delta d_H(\theta)}{57.2887}; \\ \dfrac{d\left(E_H(\theta)\right)}{dt} & = \dfrac{\partial\left(E_H(\theta)\right)}{\partial\left(d_H(\theta)\right)}\dfrac{d\left(d_H(\theta)\right)}{dt} = -\tau_i^{-1}\theta\dfrac{k_H\left[\Delta d_H(\theta)\right]^2}{57.2887}. \end{cases}$$

(3)

Obtained from the Lagrangian solution [13]， the force constant $k_H$ increases from 32 to 35 eV/Å$^2$ when the $\omega_H$ shifts from 3190 to 3260 cm$^{-1}$ at heating from 0 to 90 °C. **Figure 5**a, b show that the O:H-O bond possessing memory effect, which relaxes with momenta that depend on initial energy storage. Although passing the same temperature on the way to freezing the H-O bond at initially higher temperature remains highly active in contrast to the otherwise when they meet on the way of freeing. The enlargement of the $\Delta d_H$ by undercoordination enhances the relaxation momenta of the H-O bond in the skin.

Solving the decay function (2), yields the $\tau_i(t_i, \theta_i, \theta_f)$,

$$\tau_i = -t_i\left[Ln\left(\frac{\theta_f+b_i}{\theta_i+b_i}\right)\right]^{-1}$$

(4)

An offset of the $\theta_f$ (= 0) and the $\theta_i$ by a constant $b_i$ assures $\theta_f + b_i \geq 0$, which constraints the drain temperature $\theta_f$ in calculations ($b_i = 5$ was taken with respect to the fitting in **Figure 4**a). Matching the solution of eq (2) to the θ(t) profiles in **Figure 4**a inset, and then the experimental data of $t(\theta_i)$ in **Figure 5**c yields the respective $\tau_i$ that drops exponentially indeed with the increase of the initial temperature $\theta_i$, or with the increase of the initial energy storage and vibration frequency in **Figure 5**d, as we expected.

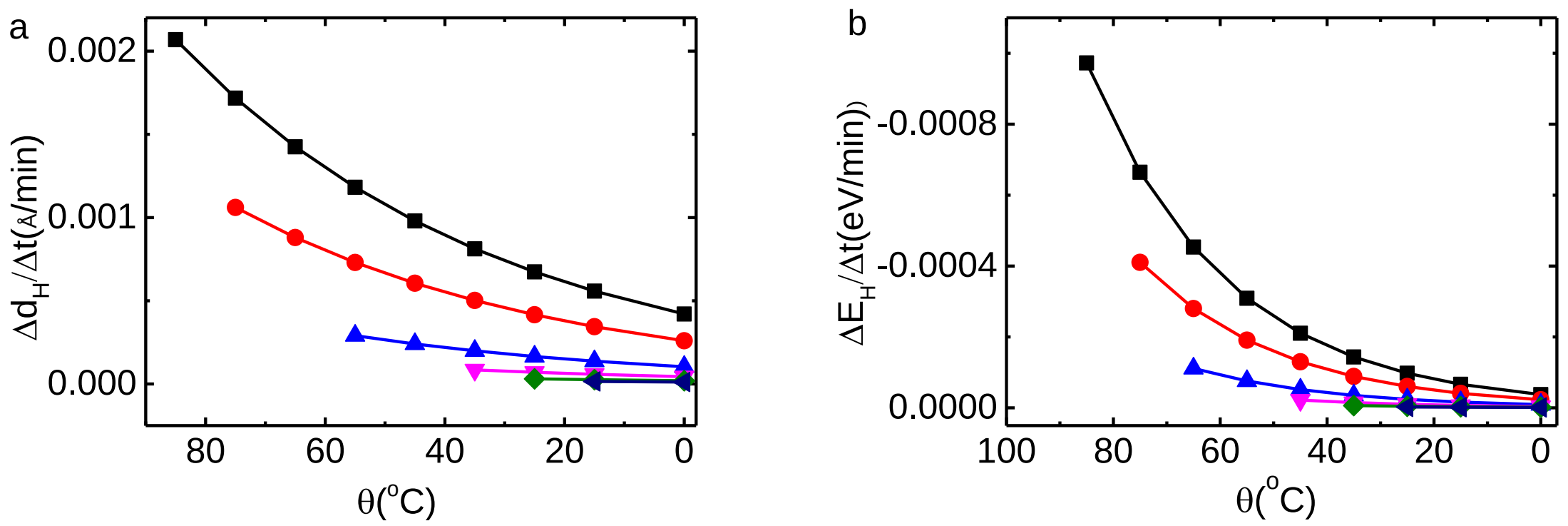

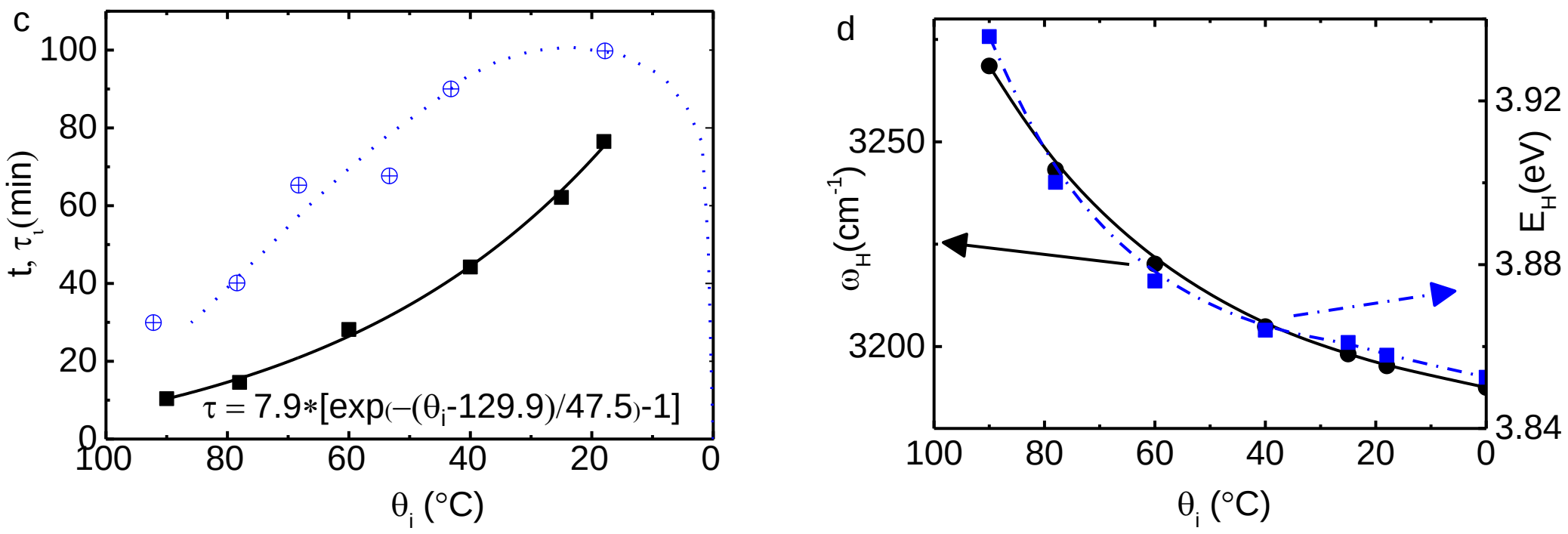

**Figure 5** **Memory effect of the H-O bond and its relaxation time.** The relaxation rates in (a) length and (b) energy of the H-O bond are higher for the one cooling from initially higher temperature than the otherwise when they meet at the same temperature on the way of freezing. Initial temperature $\theta_i$ dependence of (c) the freezing time (dotted line) (t) and the corresponding relaxation time (solid line)($\tau$), in contrast to the (d) initial $E_H(\theta_i)$ and $\omega_H(\theta_i)$[6]. The $E_H(\theta_i)$ was derived from solving Lagrangian equation[13] based on the known $\omega_H(\theta)$ and $d_H(\theta)$[2].

It is necessary to emphasize that the Mpemba effect happens only under the circumstance that the temperature drops abruptly from $\theta_i$ to the constant $\theta_f$ at the source-drain interface. Fourier solution indicates that the Mpemba crossing temperature is sensitive to the volume of the liquid (Figure 3a). If the liquid volume is too large, this effect may be prevented by heat-dissipation hindering. As experimentally confirmed by Brownridge [15], any spatial temperature decay between the source and the drain by tube sealing, oil film covering, source-drain vacuum isolating, muffin-tin like containers connecting, or even multiple sources putting into the limited volume for a refrigerator could prevent Mpemba effect from being happened.

H-O bond energy determines the critical temperature for phase transition [41]. Generally, superheating is associated with H-O bond contraction pertained to water molecules with fewer than four neighbors such as those formed the skin, monolayer film, or droplet on a hydrophobic surface with 5-10 Å air spacing between the drop and the surface presented [42]. Supercooling is associated with H-O bond elongation associated with water molecules in contact with hydrophilic surface [43] or being compressed [41]. The supercooling of the colder water in the Mpemba process [15] evidences that the initially longer H-O bond of cold water is lazier than those in the warmer water to relax at icing because of the slower momenta of relaxation.

The involvement of ionic solutes or impurities [44, 45] mediates the Coulomb coupling and the H-O bond energy because of the alternation of charge quantities and ions volumes[10]. Salting shares the same effect of heating on the H-O phonon blue shift [46-48], which is expected to enhance the velocity of heat ejection at cooling. Mass loss due to evaporation at the temporarily higher temperatures [9] affects no rate of O:H-O bond relaxation albeit the negligible amount of water to be cooled.

In summary, numerical reproduction of observations revealed the following pertaining to Mpemba paradox:

(i) O:H-O bond possesses memory effect, whose thermal relaxation defines intrinsically the rate of energy ejection. Heating stores energy to water by O:H-O bond deformation. Cooing does oppositely to emit energy with a thermal momentum that is history dependent.

(ii) Heating and skin supersolidity create gradients of thermal diffusion with a critical coefficient ratio of $\alpha_S/\alpha_B \geq \rho_B/\rho_S = 4/3$ in the liquid for heat conduction. Convection alone raises only the skin temperature.

(iii) Highly non-adiabatic ambient with step temperature change is necessary to ensure the immediately energy dissipation. The Mpemba crossing temperature is sensitive to the volume of the water being cooled. This effect will not be observable if the liquid volume is too large.

(iv) Mpemba effect takes place with a characteristic relaxation time that drops exponentially with the increase of the initial temperature or the initial energy storage of the liquid.

**Acknowledgement**

Critical reading by Yi Sun and financial support received from NSF (Nos.: 21273191, 1033003, and 90922025) China and MOE (RG29/12) Singapore are gratefully acknowledged.

**Corresponding Authors:** ecqsun@ntu.edu.sg (C.Q.).

**Affiliations:**

1. *NOVITAS, School of Electrical and Electronic Engineering, Nanyang Technological University, Singapore 639798*
2. *Center for Coordination Bond and Electronic Engineering, College of Materials Science and Engineering, China Jiliang University, Hangzhou 310018, China*
3. *Key Laboratory of Low-Dimensional Materials and Application Technologies (Ministry of Education) and Faculty of Materials, Optoelectronics and Physics, Xiangtan University, Hunan 411105, China*

**Content entry:**

**Thermal and undercoordination induced hydrogen bond relaxation defines intrinsically the ways of heat emission and heat conduction in the Mpemba paradox that happens only in the highly non-adiabatic ambient.**

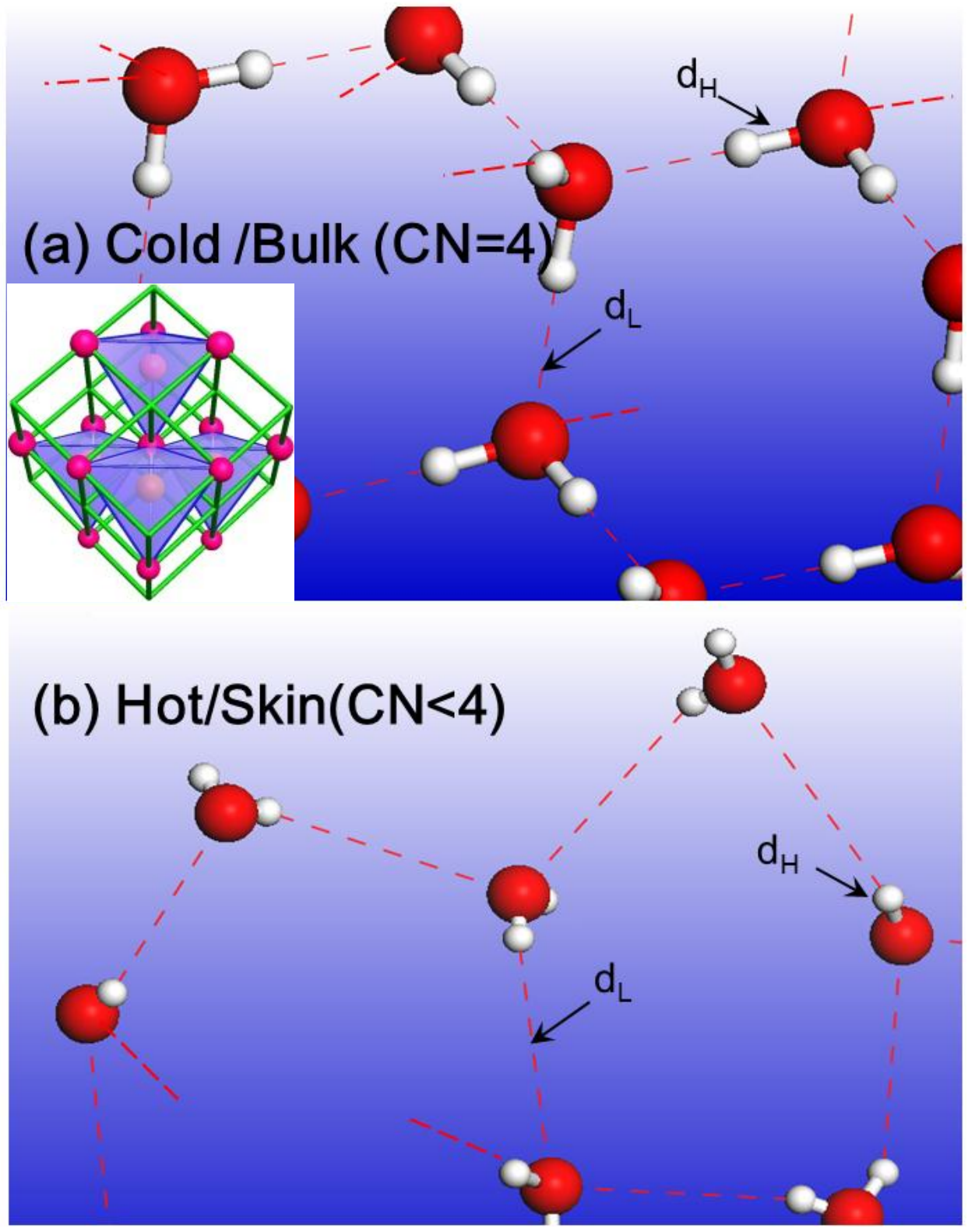

**Supplementary Information**

# 1 Joint effect of thermal and undercoordination on O:H-O bond relaxation

## *1.1 Length, phonon stiffness, and energy relaxation*

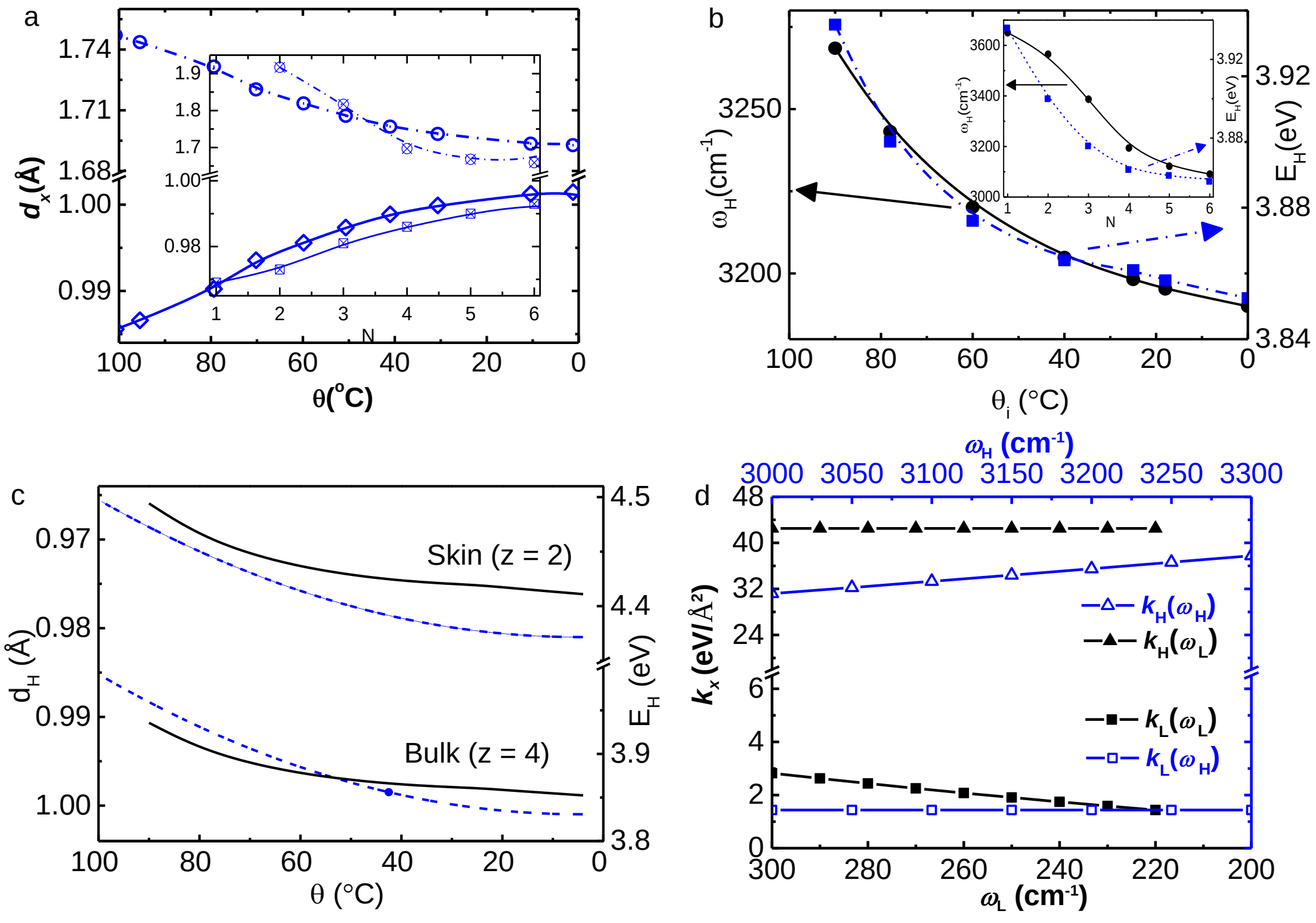

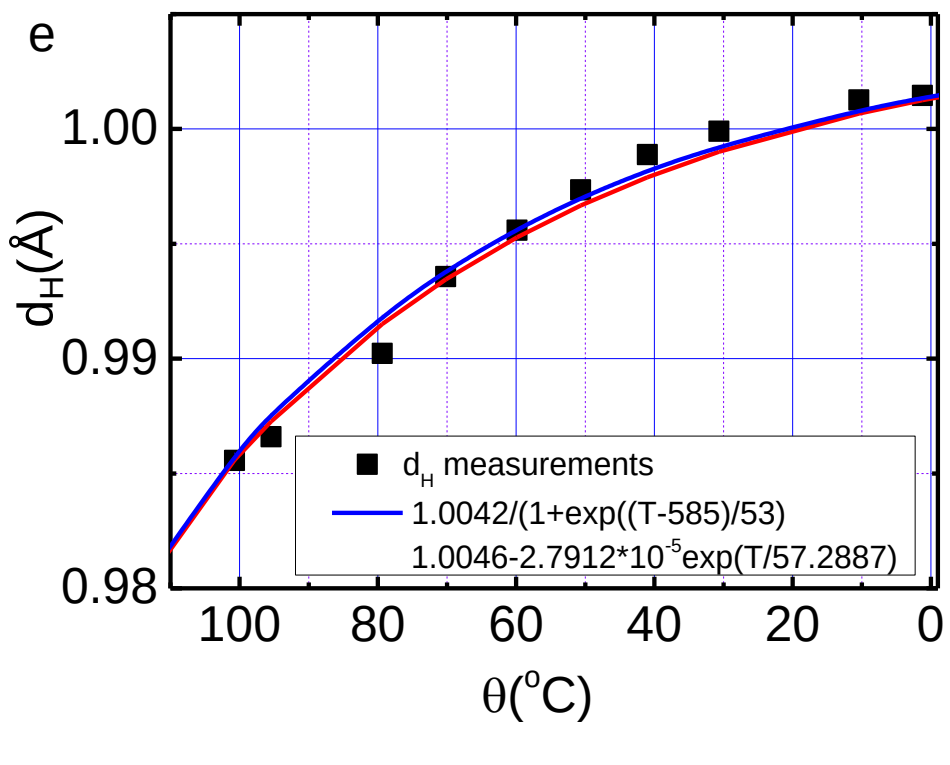

**Fig S1** (a) Profiles of thermal[1, 2] and (inset) molecular undercoordination[5] induced O:H-O bond length $d_H$ relaxation, and (b) the corresponding H-O vibration frequency $\omega_H$ and bond energy $E_H$. The $E_H$ was obtained from Lagrangian solution with the known $\omega_H$ at heating[6] and undercoordination[5-8] as input. (c) Heating and molecular undercoordination enhance each other in the $d_H$ and $E_H$ relaxation[10]. (d) The force constant $k_x(\omega_x)$ variation with Raman frequency gained by Lagrangian solution[13] with the known $\omega_x(\theta)$ and $d_H(\theta)$ as input and (e) the $d_H(\theta)$ is converted[2] from the known $\rho(\theta)$[1].

### *1.2 Skin supersolid of water ice*

As the O:H contributes only 1~2% to the system energy, one may focus on the H-O bond relaxation dynamics though this iteration applies to the H:O nonbond. The following relationship[10] couples the thermal and the undercoordination effect on the H-O bond length and energy, with the known skin $d_{OO}$ = 2.96 Å[31] (or $d_H$ = 0.84 Å)[2], shown in Fig S1c:

$$\begin{cases} d_H(z,\theta) = d_H(4,\theta) \times d_H(2,0)/d_H(4,0) \approx 0.84 d_H(4,\theta) \\ E_H(z,\theta) = E_H(4,\theta) + \Delta E_H(2,0) = 0.18 + E_H(4,\theta) \end{cases}$$

$\Delta E_H(2,0) = 0.18$ eV. Fig S1c compares the $d_H(z, \theta)$ and the $E_H(z, \theta)$ profiles for the bulk (z = 4) and the skin (z ≅ 2, N ≅ 3)[13].

Fig S2 shows the residual Raman spectra of the skin of water and ice and Table S1 the bond length, phonon stiffness, and mass density information.

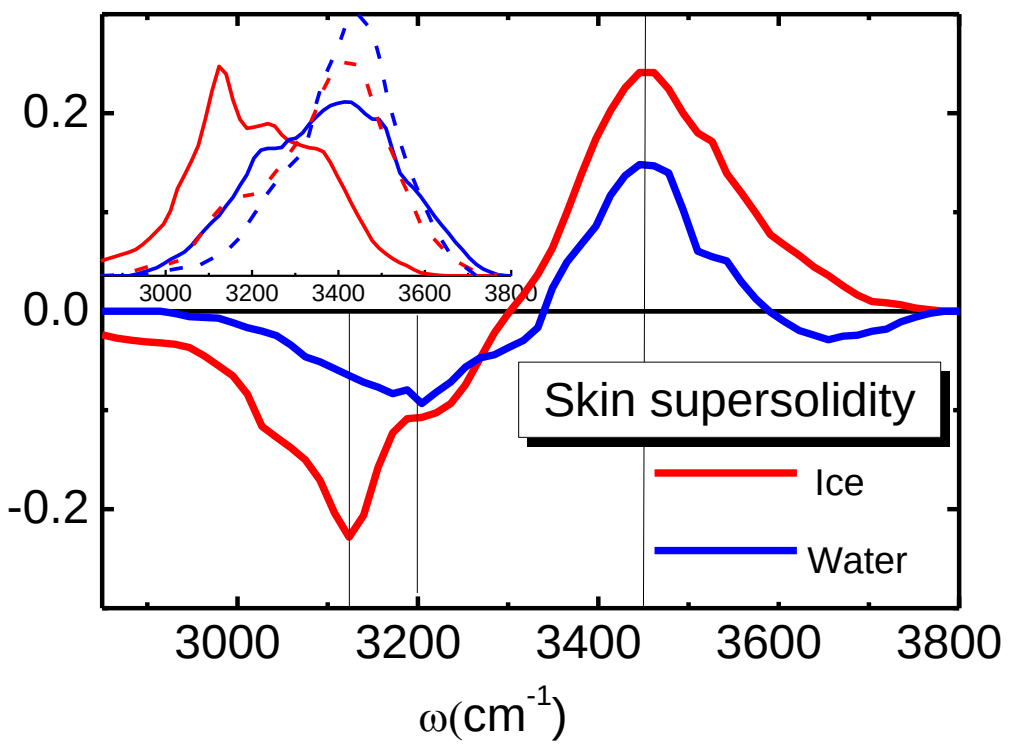

**Fig S2** Undercoordination resolved $\omega_H$ at ice and water skins. Residual Raman spectra show the components of skin and bulk of water and ice. Inset shows the raw spectra collected at difference glancing angles from ice and water [27].

**Table S1** Skin supersolidity ($\omega_x$, $d_x$, $\rho$) of water and ice derived from the measurements (indicated with refs) and methods described in [2, 13].

| | Water (298 K) | | Ice ($\rho_{min}$) | Vapor |
|---|---|---|---|---|
| | bulk | skin | bulk | dimer |
| $\omega_H$(cm$^{-1}$) | 3200[27] | 3450[27] | 3125[27] | 3650[49] |
| $\omega_L$[23] | 220 | ~180[5] | 210 | 0 |
| $d_{OO}$(Å) | 2.700[30] | 2.960[31] | 2.771 | 2.980[31] |
| $d_H$ (Å) | 0.9981 | 0.8406 | 0.9676 | 0.8030 |
| $d_L$ (Å) | 1.6969 | 2.1126 | 1.8034 | ≥ 2.177 |
| $\rho$(g·cm$^{-3}$) | 0.9945 | 0.7509 | 0.92 [50] | ≤ 0.7396 |

### *1.2 Temperature dependence of the Thermal diffusion coefficient*

The known temperature dependence of the thermal conductivity $\kappa_B(\theta)$, mass density $\rho_B(\theta)$, and specific heat under constant pressure $C_{pB}(\theta)$, given in **Fig S3**, determines the diffusion coefficient $\alpha_B(\theta)$.The skin supersolidity due to molecular undercoordination[5] changes the these quantities accordingly in the skin region. For simplicity, one can take the $\alpha_S$ in the skin (z = 2) as an adjustable parameter but it is constrained by $\alpha_S(\theta)/\alpha_B(\theta) \geq \rho_B(4°C)/\rho_S(4°C) = 4/3$ for optimization.

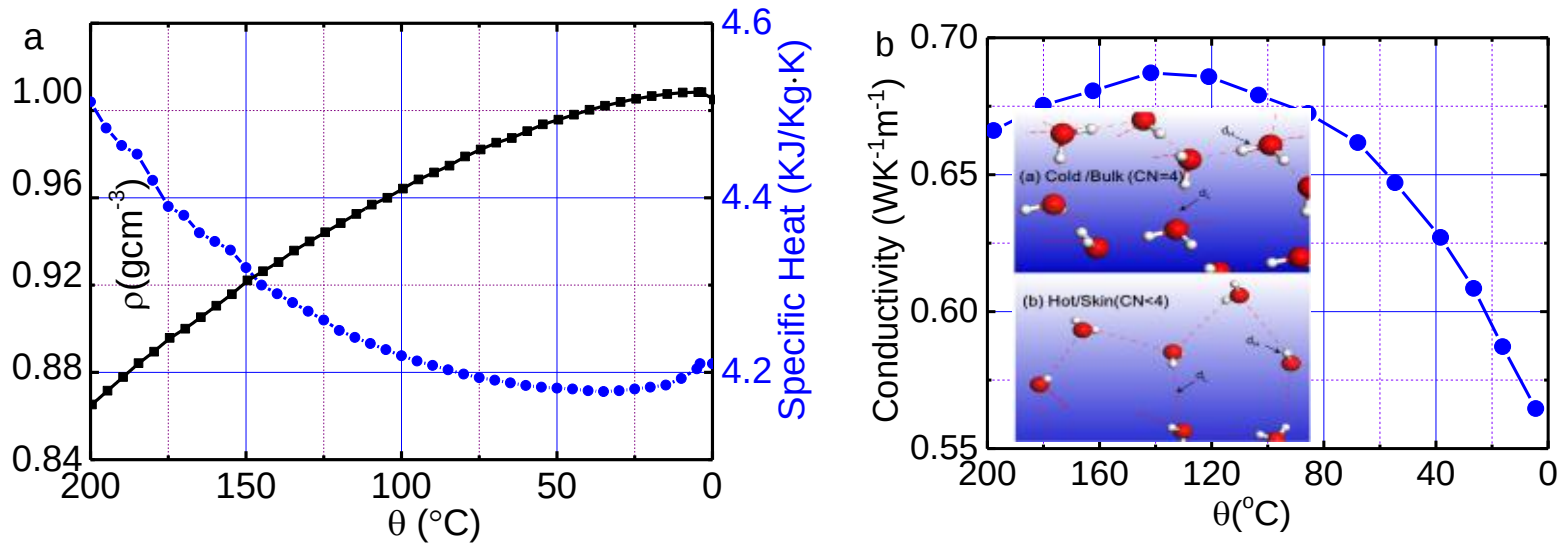

**Fig S3** Temperature dependence[1] of (a) the mass density, specific heat, (b) thermal conductivity in $H_2O$[40]. Inset in b shows that heating[23] and molecular undercoordination (CN < 4)[5] shortens the $d_H$ (molecular size) and enlarges molecular separation $d_{OO}$.